\documentclass[11pt]{article}

\usepackage{moriond,epsfig}
\usepackage{amsmath,amssymb}
\usepackage{subfigure}
\usepackage{float,multicol}

\def\ab{\bar{\alpha}}
\newcommand{\avg}[1]{\left\langle#1\right\rangle}

\begin{document}
\vspace*{4cm}
\title{Fluctuation effects in high-energy QCD}

\author{G. Soyez\footnote{On leave from the fundamental theoretical physics group of the University of Li\`ege.}}

\address{SPhT, CEA/Saclay, Orme des Merisiers, Bat 774, F-91191 Gif-sur-Yvette cedex, France}

\maketitle
\abstracts{
The recent high-energy QCD equations including the effects of fluctuations are solved numerically. We discuss their asymptotic properties and compare them with the corresponding behaviour without fluctuations as well as with the statistical-physics model obtained in the saddle point approximation. The potential consequences on phenomenology, particularly on geometric scaling violations, are drawn.
}

In our way to understand the high-energy limit of QCD, a large amount of effort is made to obtain the evolution equations including the saturation effects. Indeed, when the system becomes more and more dense, one has to consider multiple scattering and recombinations, eventually leading to saturation and unitarity. The simplest of these equations is the Balitsky-Kovchegov (BK) equation \cite{bk} which is known to admit traveling-waves solutions leading to geometric scaling $T(k^2,Y) = T(k^2/Q_s^2(Y))$, where $Q_s(Y)$ is the energy-dependent saturation scale. 

Recently, it has been proven \cite{it} that one also has to deal with fluctuation effects which become important in the dilute regime. At large $N_c$, the evolution of the amplitude takes the form of a hierarchy of equations where $T^{(k)}$ is coupled to $T^{(k)}$ through BFKL evolution, to $T^{(k+1)}$ through merging leading to saturation, and to $T^{(k-1)}$ through fluctuation effects coming from splitting.

Interestingly enough, after a coarse-graining in impact parameter, this infinite hierarchy turns out to be equivalent to a Langevin equation which is formally equivalent to the BK equation supplemented with a noise term reproducing gluon number fluctuations
\begin{equation}\label{eq:langevin}
\partial_YT(k) = \ab \chi(-\partial_{\log(k)}) T(k) - \ab T^2(k)
               + \ab \sqrt{2\kappa\alpha_s^2T(k)} \nu(k,Y),
\end{equation}
where $\ab=\alpha_sN_c/i$, $\chi(\gamma)$ is the BFKL kernel, and the noise $\nu(k,Y)$ satisfies
\[
\avg{\nu(k,Y)} = 0\qquad\text{ and }\quad
\avg{\nu(k,Y)\nu(k',Y')} = \frac{1}{\ab\pi}\delta(Y-Y')k\delta(k-k').
\]
The unknown fudge factor $\kappa$ fixes the strength of the noise and is expected to be of order 1. From this equation, one can see in which limits the different contributions to the evolution become important. On the one hand, the saturation contribution becomes relevant when $T\approx T^2$ {\em i.e.} when $T\approx 1$ which is the dense regime as expected. On the other hand, the noise term is important when $T\approx\sqrt{\alpha_s^2T(k)}$ or, equivalently, $T\approx \alpha_s^2$. This corresponds to the dilute regime where fluctuations are indeed expected to play a crucial role.

In what follows, we present the results \cite{gs} of numerical simulations of the Langevin equation \eqref{eq:langevin}. The basic idea (see \cite{levine,gs} for a detailed description) for the numerics is to introduce the probability $P(\delta Y;T_0\to T)$ that the effect of the noise changes the amplitude from $T_0$ to $T$ in a rapidity step $\delta Y$. $P$ satisfies the Focker-Planck equation
\begin{equation}
\partial_Y P(Y;T_0\to T) = \frac{\kappa\ab\alpha_s^2}{\pi} \partial_T^2 \left[ T P(Y;T_0\to T)\right],
\end{equation}
which can be solved with the initial condition $P(0;T_0\to T)=\delta(T-T_0)$ and the effect of the noise can be computed at each spatial point using the associated cumulative probability $F_{T_0,\delta Y}(T)$. We perform one step of the evolution using 
\begin{eqnarray*}
T_{\text{noise}}(k) & = & F^{-1}_{T(k,Y),\delta Y}\left\lbrack u(k)\right\rbrack,\\
T(k,Y+\delta Y) & = & T_{\text{noise}}(k,Y) + \delta Y\left\lbrack
                      \ab\chi(-\partial_L)T_{\text{noise}}(k,Y)-\ab T_{\text{noise}}^2(k,Y)
                      \right\rbrack,
\end{eqnarray*}
where $u(k)$ is a uniformly-distributed random variable between 0 and 1. For convenience, we shall use $\ab=0.2$, $L=\log(k)$ and $\tilde\kappa = \frac{10\pi}{N_c^2}\kappa$. The initial condition $T(L,Y=0)$ for the evolution is $\exp(-2L)$ when $L>0$ and $1-L$ otherwise. 

Before going into the analysis, let us note that if one performs a diffusive approximation in the BFKL kernel, Eq. \eqref{eq:langevin} simplifies to the well-known stochastic FKPP equation. This approximation is known to give relevant results for the tail of the wavefront and allows one to study analytically \cite{bd} the limit $\kappa\alpha_s^2\to 0$.

\begin{figure}
\subfigure[100 events and their average]{\includegraphics[scale=0.6]{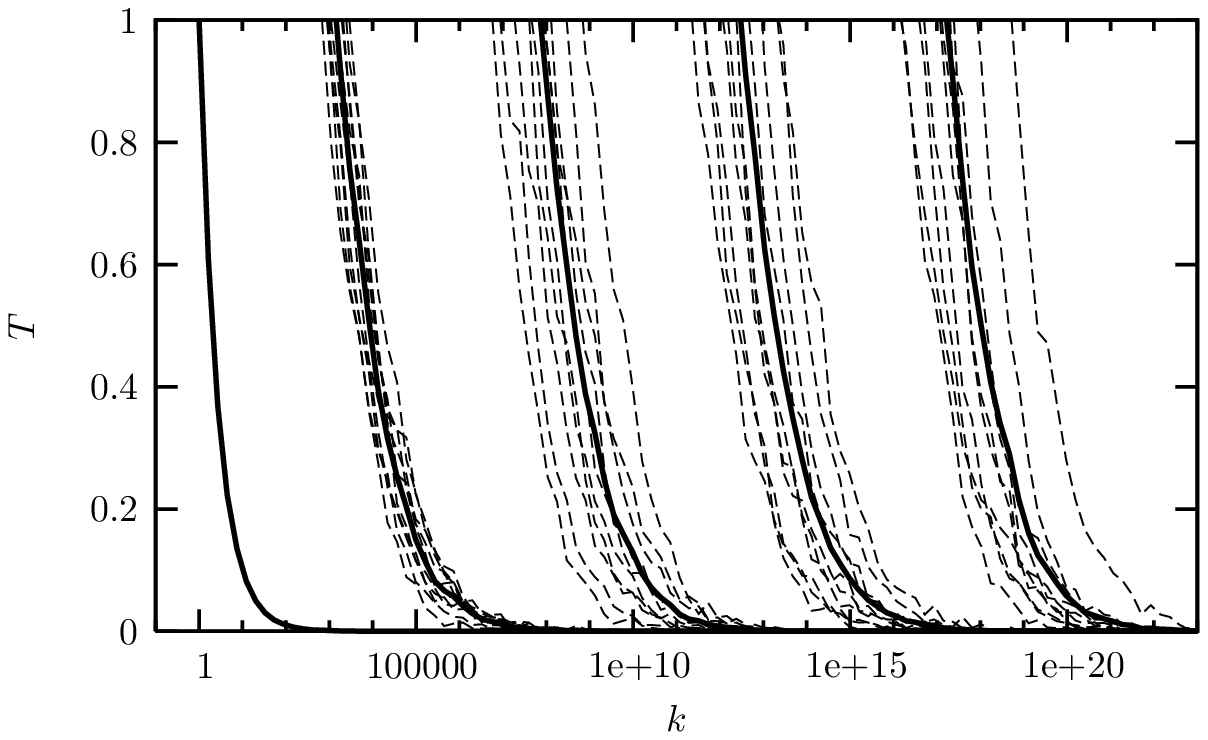}}
\subfigure[Dispersion as a function of Y]{\includegraphics[scale=0.6]{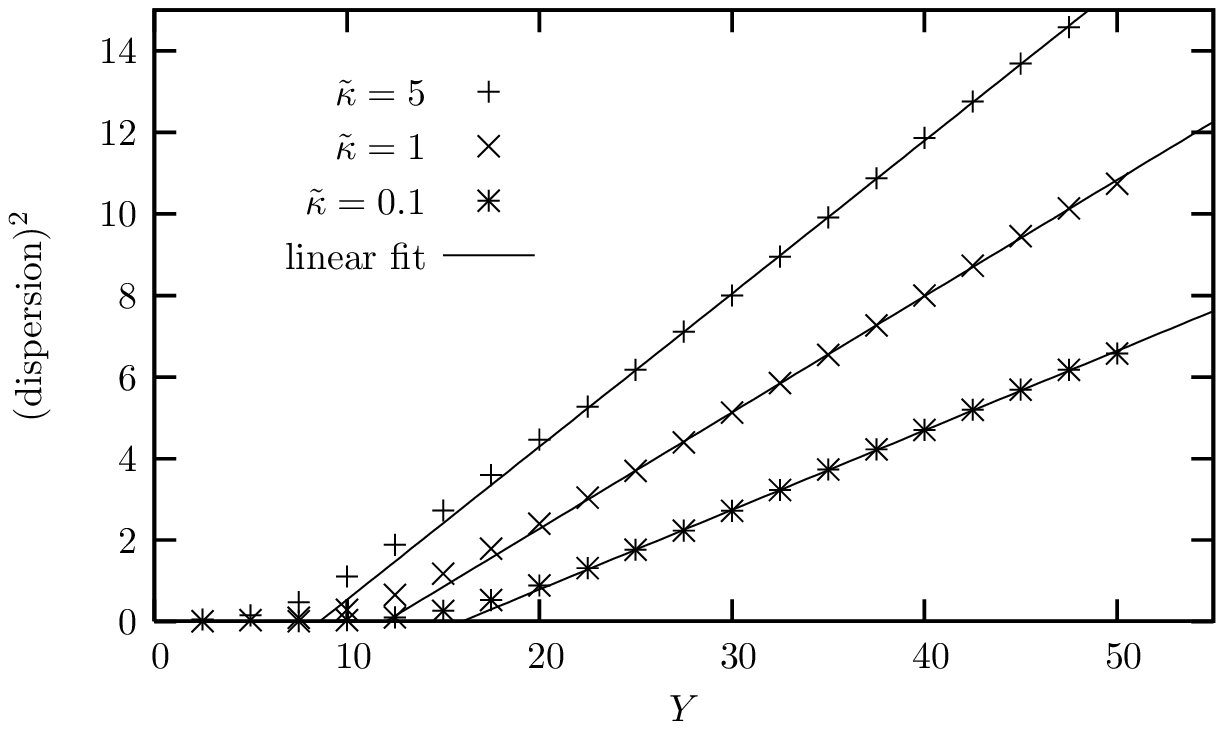}}
\caption{Results of the event-by-event study. $Y$ goes from 0 to 50.}\label{fig:disp}
\end{figure}

To begin with let us analyse a set of events generated through our method. As shown in Fig. \ref{fig:disp}, one clearly sees that each event shows a traveling-wave pattern, like in the BK case. In addition, the different events are spread. This dispersion means that although geometric scaling is preserved for each singular event, it is violated for the averaged amplitude. A more careful analysis (see the right plot of Fig. \ref{fig:disp}) shows that the dispersion, computed over 10000 events, behaves like $\sqrt{Y}$ which is typical of these random processes. From the analogy with the stochastic FKPP equation, well studied in statistical physics, we know that, for asymptotically small values of $\kappa\alpha_s^2$ the dispersion should behave like $\log^{-3}(\kappa\alpha_s^2)$. However, in order to observe that behaviour, one needs values of $\kappa\alpha_s^2$ as small as $A0^{-20}$ which cannot be reached by our method (see \cite{egm} for a model), 
It is important to note that, in the early stages of the evolution, we do not observe much dispersion, which may explain the fact that geometric scaling is observed in the HERA data. 

\begin{figure}[H]
\includegraphics[scale=0.58]{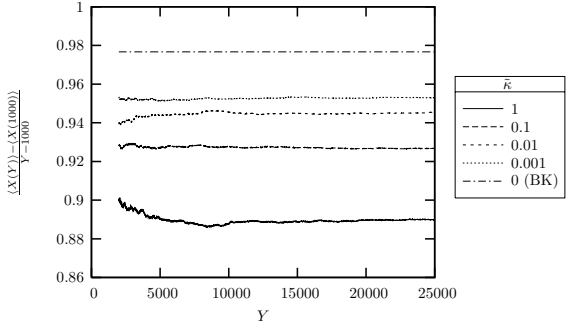}
\caption{Asymptotic speed of the traveling wave for different strengths of the noise, compared with the BK asymptotic speed.}\label{fig:speed}
\end{figure}

Another property that can be studied is the asymptotic speed of the traveling wave which is related to the saturation scale $Q_s^2(Y)=k_0^
2\exp(vY)$. For asymptotically small $\kappa\alpha_s^2$, should decrease w.r.t. BK according to \cite{bd}
\[
v^* \underset{\alpha_s^2\kappa\to 0}{\to} v_c^{(BK)} - \frac{\ab\pi^2\gamma_c\chi''(\gamma_c)}{2\log^2(\alpha_s^2\kappa)}.
\]
Although we have not studied this asymptotic regime, we confirm this slowing down \footnote{The asymptotic formula predicts a negative speed for $\alpha_s^2\kappa\sim 1$ which proves the importance of subleading corrections.} as shown on Fig. \ref{fig:speed} for realistic values of the noise strength. Note also that one has
 to evolve up to unphysical values of $Y$ to observe the asymptotic velocity.

\begin{multicols}{2}

\begin{figure}[H]
\includegraphics[scale=0.6]{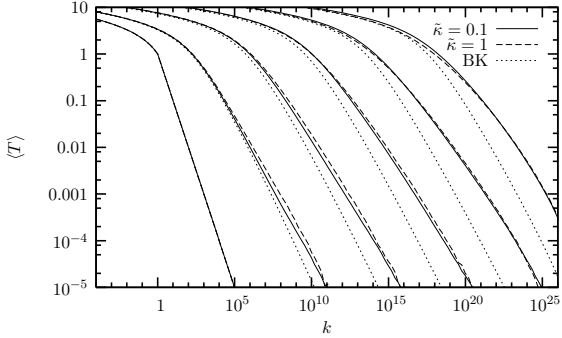}
\caption{Evolution of the average amplitude $\avg{T}$ with rapidity $Y=0,10,...,50$. One can see the effect of the noise by comparison with the BK curve.}\label{fig:front}
\end{figure}

Let us now proceed with the study the properties of the average amplitude, obtained from a set of 10000 events. The effect of the noise w.r.t. BK evolution without fluctuation is manifest in Fig. \ref{fig:front}. As rapidity increases, the scaling violations become more and more important and the average front has a smaller slope. One also remarks that, although the asymptotic speed is smaller than for BK, in the early stages of the evolution, this speed appears to be larger. For values of $Y\approx 40$, one sees however from the dashed curve in Fig. \ref{fig:front} that this speed is already below the BK speed.

\begin{figure}
\subfigure[Form of the wavefront ($Y=20$)]{\includegraphics[scale=0.6]{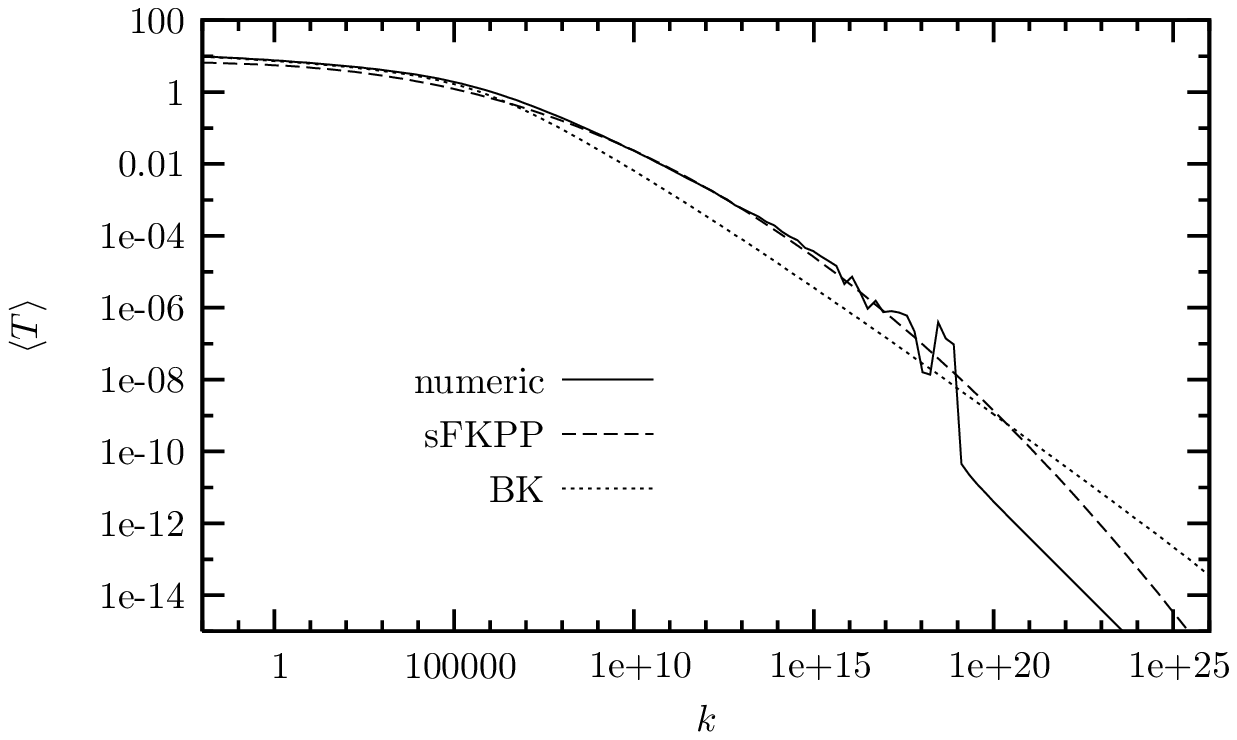}}
\subfigure[Numerical results for $\avg{T^2}$.]{\includegraphics[scale=0.6]{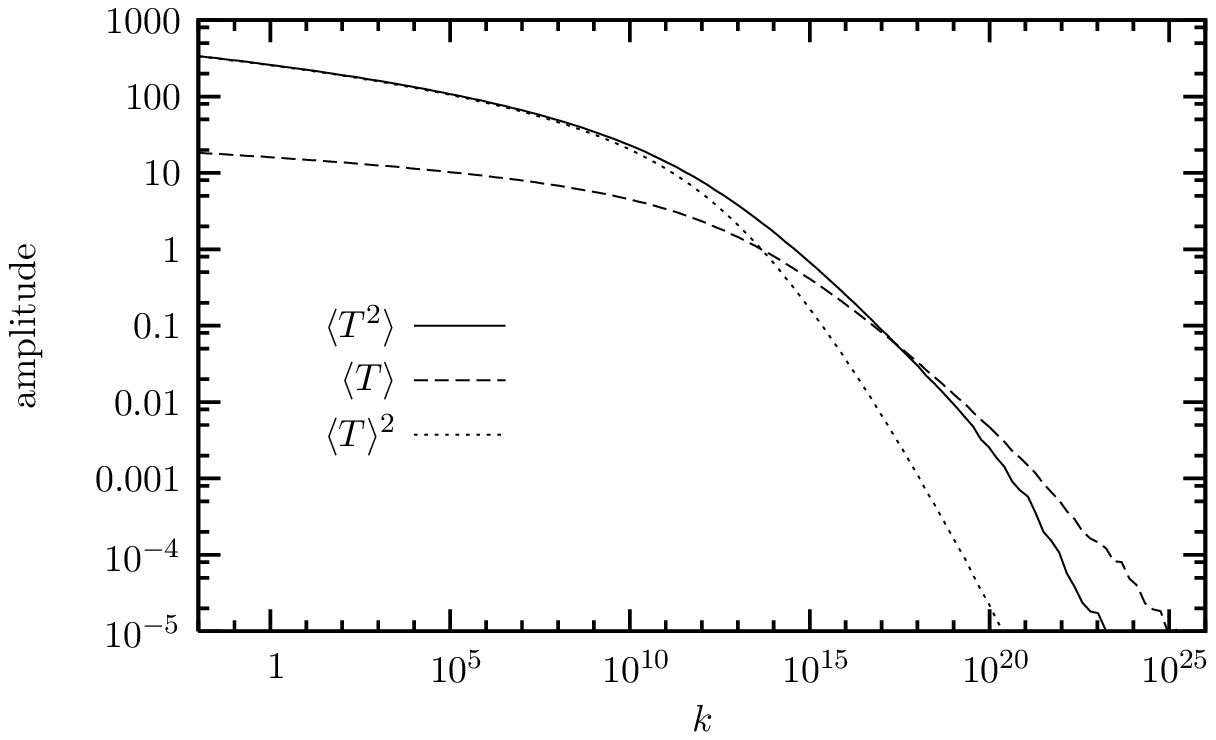}}
\caption{Comparison of the numerical simulations ($\tilde\kappa=1$) with the stochastic FKPP predictions. Left plot: form of the wavefront. Right plot: correlations $\avg{T^2}$ together with $\avg{T}^2$ and $\avg{T}$ for comparison.}\label{fig:sfkpp}
\end{figure}

Finally, we have compared the results of our simulations with the predictions of the stochastic FKPP equation, as shown in Fig. \ref{fig:sfkpp}. On the left plot, presenting the results for $\avg{T}$ at $Y=20$, it is obvious that the diffusive approximation (stochastic FKPP model) agrees very well with the numerical simulations \footnote{The discrepancy at $T\gtrsim 10^{-9}$ is a numerical artefact.}. The scaling violations for $10^{-4}\lesssim T \lesssim 1$ are well observed and the rapid fall-off seen in the far tail of the front is coming from the fact that each event has a compact front. Theoretically, it can be seen from the probability $P$ and, in physical terms, it means that the effect of the fluctuations is to cut off the ultraviolet tail.
\end{multicols}

From a set of events generated using Eq. \eqref{eq:langevin}, we can also generated higher-order correlators. The diffusive approximation predicts
\[
\avg{T}\approx \avg{T^2} \approx \dots \approx \langle T^k\rangle \approx \dots
\]
for asymptotic values of $Y$. As shown on Fig. \ref{fig:sfkpp} (right plot), numerical simulations qualitatively agree with this result for the tail of the front although the asymptotic regime is not yet fully reached. In the saturated domain, the diffusive approximation is not sufficient since it misses the logarithmic behaviour in the infrared. However, this region is characterised by $\avg{T^2}\approx \avg{T}^2$ as confirmed by simulations.

It is interesting to notice that the sFKPP model agrees already at values of $Y$ which are not extremely high as compared to the time needed to reach the asymptotic speed. Actually, it has been proven \cite{imm}, in the asymptotically-small $\kappa\alpha_s^2$ limit, that the formation time for the wavefront is of order $\beta^{-1}\gamma_c^{-2}\log^2(\kappa \alpha_s^2)$ with $\beta\approx 48.2$. This is faster than the equivalent time for the BK equation and might have consequences on geometric scaling. Although this may depend on our choice of initial condition, it seems that one has a competition between a short time needed to reach the asymptotic form of the wavefront and the scaling violations induced by dispersion which (see Fig. \ref{fig:disp}) are not observed in the early stages of the evolution. This may leave a region in rapidity in which geometric scaling is verified.

To summarise, we have seen from numerical simulation of the Langevin equation \eqref{eq:langevin} that the contribution from fluctuations have a large impact on the high-energy evolution in QCD. It modifies the rapidity-dependence of the saturation scale and introduces geometric scaling violations. Our results, obtained for physical values of the coupling, are in agreement with the predictions coming from the stochastic FKPP equation. 

In the future, further analysis should be performed in order to clarify the question of geometric scaling in the experimentally accessible range of rapidity where preasymptotic effects cannot be neglected. Furthermore, an extension of the solution to the complete hierarchy of equations, including the impact-parameter dependence, would be a precious part of our understanding of QCD at high-energy.

\section*{Acknowledgements}
G.S. is funded by the National Funds for Scientific Research (Belgium). I would like to thank R.~Peschanshi and E.~Iancu for carefully proofreading these proceedings.

\end{document}